# Evidence for early identification of Alzheimer's disease

**Athanasios Alexiou, Panayiotis Vlamos[1]**

**Abstract**

Alzheimer's disease is a human brain disease that affects a significant fraction of the population by causing problems with short-term memory, thinking, spatial orientation and behavior, memory loss and other intellectual abilities. Up to date there is no singular test that can definitively diagnose Alzheimer's disease, although imaging technology designed to detect Alzheimer's plaques and tangles is rapidly becoming more powerful and precise. In this paper we introduce a decision-making model, based on the combination of mitochondrial hypothesis-dynamics with the role of electromagnetic influences of the metal ions into the inner mitochondrial membrane and the quantitative analysis of mitochondrial population. While there are few disappointing clinical-trial results for drug treatments in patients with Alzheimer's disease, scientific community need alternative diagnostic tools rather investing mainly in amyloid-targeting drugs.

**Keywords:** Alzheimer's Disease (AD), Mitochondrial Electrophysiology (ME), Mild Cognitive Impairment (MCI), Atomic Force Microscopy (AFM)

## 1. Strategies for early diagnosis

There are few cases where scientists and pharmaceutical companies undergo trials in cognitively normal individuals and patients with mild cognitive impairment (MCI) combining already approved therapeutic agents with neuropsychological tests, blood and urine analyses, quantitative magnetic resonance imaging (MRI) or magnetic resonance spectroscopy (MRS), in order to evaluate the current hypotheses on the etiology of AD. Additionally, there are several tests in the market for the early diagnosis of MCI and AD, such as the 30-minute cognition test, which involves recording of brainwaves while patient listen to auditory stimuli. In this 30-minute cognition test for example, researchers found that AD patients process the sounds differently than healthy adults. Additionally, we can refer to nuclear medicine techniques that are established for diagnostic and research purposes such as Positron Emission Tomography (PET) and CAT/CT and Nuclear Magnetic Resonance Imaging (NMRI/ MRI), even though there are major limitations in terms of either cost or image resolution, as well as patient irradiation or Fluorescence spectroscopic techniques that are capable of single cell or molecule detection and imaging. These novel techniques, combined with the capabilities of several Nanotechnological techniques, such as $\mu$MRI and Scanning Tunneling Microscopy (STM), may offer the early detection of AD in terms of modern biophysics. According to several studies, neurodegenerative diseases are highly associated to mitochondrial disorders i.e. the low production of ATP, decreased mitochondrial membrane potential and mutations in mtDNA are factors related to disorders like AD.

[1] Correspondence should be addressed to Dr. Panayiotis Vlamos, Associate Professor in the Ionian University, Department of Informatics, Bioinformatics and Human Electrophysiology Laboratory, Plateia Tsirigoti 7, 49100 Corfu, Greece, vlamos@ionio.gr

In the brain, mitochondrial function declines with age and this functional decline associates with increased mitochondrial biogenesis. In various neurodegenerative disease states brain mitochondrial function declines even further, perhaps to the point where mitochondrial biogenesis can no longer compensate for functional declines (Swerdlow et al., 2004; Swerdlow et al., 2009; Onyango et al., 2010). While AD can be genetically classified as familiar or sporadic, researchers proposed that beside the main components of the plaques and tangles, amyloid-β (Aβ) and Tau proteins, the case of sporadic AD is also a consequence of decreased mitochondrial function over age. The overexpression of Aβ causes an alteration in the mitochondrial fission and fusion proteins resulting in mitochondrial dysfunction, mitochondrial fragmentation, increase in reactive oxygen species (ROS) and ATP production and reduced mitochondrial membrane potential (Wang et al., 2008). Following a modern biophysical explanation of abnormalities in the electron flow through the inner mitochondrial membrane, defined as 'electric thrombosis' (Alexiou et al., 2011), we examined whether these unusual interactions and electrical complexes are symptoms and etiology of AD or just a consequence of the disease. In this paper we propose the notion Mitochondrial Electrophysiology (ME) as a valuable novel method for use by clinical investigators to study brain activity in patients with neurological illness. While the role of several metal ions is already known, Copper, Zinc and Iron metabolisms seem to play fundamental role in the mitochondrial dynamics and the quality of mitochondrial population. Therefore, the proposed procedure combines peripheral electrophysiology measurements with identification of electromagnetic interaction in subcellular level and offers a comprehensive statistical evaluation of these factors and AD progress. Our decision-making model, named AVAD, concerns physical and cellular-subcellular evaluation such as reported medical history and nutritionals habits, quantitative analysis and distribution of mitochondrial dynamics, electrophysiology measurements, *mt*DNA analysis and metal ions metabolism.

## 2. Mitochondrial dynamics and amyloid-β interaction

Recent studies have shown that mitochondria are significantly reduced in AD, supporting a topographic and probably temporal relationship between neuronal oxidative damage and mitochondrial abnormalities (Hirai et al., 2001). Although the pathological mechanism for AD is still unknown, the predominant hypothesis is that excess Aβ production results in cellular toxicity. Transgenic mouse models over-expressing amyloid precursor protein (APP) lead to amyloid plaques associated with activation of inflammatory cells, localized loss of neurons, and some cognitive behavioral changes (Dodart et al., 2005). Ab can localize to mitochondria, and this interaction has been suggested to underlie in part their cytotoxic effects (Manczak et al., 2006; Lustbader et al., 2004).

### 2.1 The mitochondrial hypothesis in AD

The mechanisms that cause the profound degeneration and loss of neurons in AD are not known, and existing information is incomplete. Abnormal processing or modification of APP and the cytoskeletal protein tau are involved in the pathogenesis resulting in amyloid (Aβ) deposits and neurofibrillary changes consisting of paired helical filaments, NFTs and dystrophic neurites (Hardy et al., 2002).

Cortical and hippocampal neuronal degeneration could be the consequence of a combination of several mechanisms including perturbations in protein metabolism, excitotoxicity, oxidative stress, mitochondrial perturbations, and inflammation. The possible specific mechanisms for neuronal degeneration in AD may involve dysfunction of NMDA receptors (Sze et al., 2001; Kemp et al., 2002), deregulation of $Ca^{2+}$ and mitochondrial homeostasis (Mattson et al., 1993; Reddy et al., 2008), defects in synapses (DeKosky et al., 1990; Terry et al., 1991; Martin et al., 1994; Sze et al., 1997; Selkoe, 2002), abnormalities in the metabolism of APP and presenilin proteins, toxic actions of A protein derived from APP (Yankner et al., 1989; Younkin, 1995), and cytoskeletal pathology (Fath et al., 2002; Rapoport et al., 2002). There are possible disease links between intraneuronal A and mitochondria suggesting an intracellular toxicity of A (Anandatheerthavarada et al., 2003; Devi et al., 2006; Manczak et al., 2006). Human AD autopsy brain shows evidence for mitochondrial impairments (Reddy et al., 2008). High mitochondrial APP levels mirror abnormalities in respiratory chain subunit levels and activity and enhanced ROS production (Devi et al., 2006). A can interact with the mitochondrial matrix protein A -binding alcohol dehydrogenase in human AD brain and is believed to participate in mitochondrial dysfunction and oxidative stress (Lustbader et al., 2004), while increased production of A may be a consequence of neuronal apoptosis. Additionally, defects in mitochondrial dynamics, fusion and fission have been shown to decrease mitochondrial movement. Empirically, fusion-deficient mitochondria display loss of directed movement, following Brownian motion without exact formulation of movement properties (Chen et al., 2003). In neurons lacking mitochondrial fusion, both increased mitochondrial diameter due to swelling and aggregations of mitochondria seem to block efficient entry into neurites, resulting in a death of mitochondria in axons and dendrites (Chen et al., 2007). These defects result in improperly developed neurons or gradual neurodegeneration. It is obvious that neurodegeneration that occurs in AD has not been cleared identified and even more many risk factors such as aging, increased oxidative stress, inflammatory processes, mitochondrial dysfunction, alterations in mitochondrial morphology and even cell cycle abnormalities have been proposed as components of AD pathogenesis.

**2.2 Amyloid- interacting metal ions. The case of imaging diagnosis**

Metal ions are known to play an important role in many neurodegenerative diseases beside AD including, aceruloplasminemia, amyotrophic lateral sclerosis, Huntington disease, Menkes disease, occipital horn syndrome, Parkinson disease, prion disease, Wilson disease etc. In these diseases improper regulation of redox active metal ions can induce oxidative stress by producing cytotoxic ROS. For example, Copper binding proteins play important roles in the establishment and maintenance of metal-ion homeostasis, in deficiency disorders with neurological symptoms like Menkes disease and Wilson disease and in neurodegenerative diseases like AD. The interaction of Fe and Cu with A may lead to production of $H_2O_2$ due to double electron transfer to $O_2$ (Bush, 2008) resulting in differentiations from the normal measurements. The production of $H_2O_2$ is known to be partly responsible for oxidative injury in cases of AD (Nazem et al., 2011). Furthermore, the interaction of A with cell membrane is enhanced by Zn and Cu (Cui et al., 2005) and Cu could play an important role in the neurotoxicity of A (Huang et al., 1999).

It is obvious that imaging of these ion metals is very important in order to understand the role of metals in neurodegenerative disorders in general. Imaging techniques like X-ray fluorescence microscopy (XFM), particle induced X-ray emission (PIXE), energy dispersive X-ray spectroscopy (EDS), laser ablation inductively coupled mass spectrometry (LA-ICP-MS), secondary ion mass spectrometry (SIMS) or magnetic resonance imaging (MRI) are widely researched for the imaging of metals in intact biological cells and tissues with high spatial resolution and detection sensitivity (Skaat et al., 2009; Hofmann-Amtenbrink et al., 2010; Nazem et al., 2011). However, these techniques are mostly useful for AD verification rather than early diagnosis since the amyloid plaques formation appears in a later stages.

### 3. The AVAD decision-making model

The proposed novel procedure for early identification of AD aims at combining bioinformatics methods with imaging techniques and electrophysiology measurements with associated known mitochondrial markers in order to identify symptomatology of AD in early stages. The AVAD model consists of the statistical evaluation of the four different clinical assessments below, associated with the mitochondrial hypothesis and their identified correlation with progressive AD.

**3.1 Evaluation of mitochondrial population as a dynamic system**

We established a new dynamic mathematical model for the association of alterations in mitochondrial distribution and morphology due to imbalance motility. This new stochastic model replaces the general hypothesis of Brownian motion and enables a more accurate detection of movement's dysfunction as it has been cleared identified in our simulation process. While the motility plays an important role on fission and fusion, in the case of decreased functionality creates clusters of mitochondria or even more isolated individual mitochondria as this was demonstrated in cases of skin biopsies.

**3.2 Electromagnetic interactions in the mitochondrial population**

It is cleared form the simulation results and the microscopy testing, that there is a great impact form the reflection of decreased mitochondrial motility on       of inner membrane. The pathogenesis on inner membrane potential had been investigated through a total new experimental approach on the way that metal ions Cu and Fe interact to each other and with oxygen and amyloid- peptide (A ) as well, in the specific biological environment. This study is an extension of the initial formulation of the electric complexes that occurs due to the phenomenon of 'electric thrombosis' in regards to certain T$c$ (temperature) and pH levels, in order to identify the existence of $H_2O_2$ (hydrogen peroxide) or $OH^*$ (hydroxyl radical).

**3.3 Metal ions metabolim and mtDNA analysis**

A detailed statistical analysis took place in order to identify all the correlations between Copper metabolisms, cytotoxic effects of Zn and existence of heavy metals in combination with clinical assessment for early diagnosis of MCI.

For this assessment, nutritional habits of the population sample were taken also into account as well as weight maintenance and physical activity. Additionally increased levels of mtDNA mutations and alterations in complex IV subunits analysis were also evaluated in patients.

**3.4 Frequencies and electromagnetic analysis**

EEG frequencies, bio potential measurements and electromagnetic influences were taken also in consideration; monitoring cases of decreased 'natural superconduction' of inner membrane and reduction of ATP in the mitochondrial population using AFM, were statistical evaluated, as evidences for MCI cases.

**3.5 Clinical Study**

We used clinical assessments of an 86 control asymptomatic population in order to test a sample of 45 MCI subjects and a sample of 25 mild to moderate AD patients. The subjects were for 60 to 85 years old. The groups were also compared and evaluated in terms of demographic, history health, cardiovascular risk factors, chronic diseases and identification of genetic disorders mainly of copper metabolism leading to severe copper toxicity. Copper is known to interact with amyloid precursor protein and -amyloid peptide in the self-aggregating plaques and neurofibrillary tangles characteristic of AD and may contribute to the pathogenesis of this disorder via cellular oxidative stress (Madsen et al., 2007; Zucconi et al., 2007; Lutsenko et al., 2007; Taylor et al., 2002; Dingwall, 2007; Ha et al., 2007). Brain normally contains a certain trace concentration of metal ions like Zinc, Copper and Iron, which possess different physiological roles (Bush, 2008). Any disruption in the homeostasis of these trace elements leading to higher concentrations of these ions may impose several proteins and membrane lipids to toxic effects of these trace elements and finally end in production of ROS (Bush, 2008; Castellani et al., 2007; Smith, 2006; Huang et al., 2004). Skin biopsies results were used for immunochemistry and electron micrographs on Schwann cells and dermal nerves and were evaluated using multinomial regression analysis. EEG measurements and skin bio potentials were also evaluated using Fisher's exact test and Chi-square test. AFM was used in order to confirm measurements on the natural superconductivity on inner mitochondrial membrane and evaluate surface structures in the cases of mitochondria interaction and inner membrane potential distribution. Mitochondria size and count were also evaluated and measured and the statistical significance to the control group was assessed using Student's t-test. In the overall case, the sensitivity and specificity of the protocol according to the specific combined clinical assessments was calculated on 95.6%, a secure rate for the early identification of mild AD symptoms.

4. **Future applications**

In our case in question, we evaluated bio potential measurements in combination with AFM study and EEG frequencies, in order to identify mitochondrial dysfunction and statistical evaluate a novel decision-making model in targeted sample population, concerning MCI and early stages of AD.

The modern biophysics seems to offer powerful tools that are not yet been applied in bioengineering and molecular biology. This novel model is innovative and accurate but mostly has a low cost against imaging techniques and can be designed in the future for individualized diagnosis as a mobile device. Latest news from the drug developers' clinical trials in AD patients is quite pessimistic in order to prove amyloid hypothesis as a diagnostic symptom and effective treatment (Callaway, 2012). Nevertheless the exclusively targeting of amyloid-, alongside the tardiness disease progression and drug application are disappointing indicating an ineffective strategy. This new hybrid decision-making model combines bioengineering with modern biophysics. Definitely we do not propose a drug treatment but a more sophisticated complex tool for evidence's assessment on identification of AD.

**References**

[1]. Alexiou A, Rekkas J, Vlamos P (2011) Modeling the mitochondrial dysfunction in neurogenerative diseases due to high H+ concentration. Bioinformation 6(5):173-175

[2]. Anandatheerthavarada HK, Biswas G, Robin MA, Avadhani NG (2003) Mitochondrial targeting and a novel transmembrane arrest of Alzheimer's amyloid precursor protein impairs mitochondrial function in neuronal cells J. Cell Biol. 161:41-54.

[3]. Bush, AI. (2008) Drug development based on the metals hypothesis of Alzheimer's disease. J Alzheimer's Dis. 15(2):223-40.

[4]. Callaway, E. (2012) Alzheimer's drugs take a new tack. Nature 489:13–14 doi:10.1038/489013a

[5]. Castellani, RJ., Moreira, PI., Liu, G., Dobson, J., Perry, G., Smith, MA., et al. (2007) Iron: the Redox-active center of oxidative stress in Alzheimer disease. Neurochem Res. 32(10):1640-5.

[6]. Chen H, Chan DC (2009) Mitochondrial dynamics–fusion, fission, movement, and mitophagy– in neurodegenerative diseases, Human Molecular Genetics 18(2):R169–R176

[7]. Chen, H., Detmer, S.A., Ewald, A.J., Griffin, E.E., Fraser, S.E. and Chan, D.C. (2003) Mitofusins Mfn1 and Mfn2 coordinately regulate mitochondrial fusion and are essential for embryonic development. J. Cell Biol., 160, 189–200.

[8]. Chen, H., McCaffery, J.M. and Chan, D.C. (2007) Mitochondrial fusion protects against neurodegeneration in the cerebellum. Cell, 130, 548–562.

[9]. Cui, Z., Lockman, PR., Atwood, CS., Hsu, CH., Gupte, A., Allen, DD., et al. (2005) Novel D-penicillamine carrying nanoparticles for metal chelation therapy in Alzheimer's and other CNS diseases. Eur J Pharm Biopharm. 59(2):263-72.

[10]. DeKosky ST, Scheff SW (1990) Synapse loss in frontal cortex biopsies in Alzheimer's disease: correlation with cognitive severity Ann. Neurol. 27:457-464.

[11]. Devi L, Prabhu BM, Galati DF, Avadhani NG, Anandatheerthavarada HK (2006) Accumulation of amyloid precursor protein in the mitochondrial import channels of human Alzheimer's disease brain is associated with mitochondrial dysfunction J. Neurosci. 26:9057-9068.

[12]. Dingwall, C. (2007) A copper-binding site in the cytoplasmic domain or BACE1 identifies a possible link to metal homeostasis and oxidative stress in Alzheimer's disease. Biochem Soc Trans. 35:571–3.

[13]. Dodart JC, May P (2005) Overview on rodent models of Alzheimer's disease. Curr Protoc Neuroscience, Chapter 9, Unit 9.22.

[14]. Fath T, Eidenmuller J, Brandt R (2002) Tau-mediated cytotoxicity in a pseudo hyperphosphorylation model of Alzheimer's disease J. Neurosci. 22:9733-9741.

[15]. Ha, C., Ryu, J., Park, CB. (2007) Metal ions differentially influence the aggregation of deposition of Alzheimer's -amyloid on a solid template. Biochemistry. 46:6118–25.

[16]. Hardy J, Selkoe DJ (2002) The amyloid hypothesis of Alzheimer's disease: progress and problems on the road to therapeutics. Science 297(5580):353-356.

[17]. Hirai K, Aliev G, Nunomura A, Fujioka H, Russell RL, Atwood CS, Johnson AB, Kress Y, Vinters HV, Tabaton M, Shimohama S, Cash AD, Siedlak SL, Harris PL, Jones PK, Petersen RB, Perry G, Smith MA (2001) Mitochondrial abnormalities in Alzheimer's disease. J Neuroscience 21(9):3017-23.

[18]. Hofmann-Amtenbrink, M., Hofmann, H., Montet, X. (2010) Superparamagnetic nanoparticles - a tool for early diagnostics. Swiss Med Wkly 140(w13081).

[19]. Huang, X., Cuajungco, MP., Atwood, CS., Hartshorn, MA., Tyndall, J., Hanson, GR., et al. (1999) Cu(II) potentiation of Alzheimer Abeta neurotoxicity: correlation with cell-free hydrogen peroxide production and metal reduction. J Biol Chem. 274:37111-6.

[20]. Huang, X., Moir, RD., Tanzi, RE., Bush, AI., Rogers, JT. (2004) Redox-active metals, oxidative stress, and Alzheimer's disease pathology. Annals New York Acad. Sci. 1012, 153-163.

[21]. Kemp JA, McKernan RM (2002) NMDA receptor pathways as drug targets Nat. Neurosci. (suppl) 5:1039-1042.

[22]. Lustbader JW, Cirilli M, Lin C, Xu HW, Takuma K, Wang N, Caspersen C, Chen X, Pollak S, Chaney M, Trinchese F, Liu S, Gunn-Moore F, Lue L-F, Walker DG, Kuppsamy P, Zewier ZL, Arancio O, Stern D, Yan SS, Wu H (2004) ABAD directly links A to mitochondrial toxicity in Alzheimer's disease Science 304:448-452.

[23]. Lutsenko, S., Barnes, NL., Bartee, MY., Dmitriev, OY. (2007) Function and regulation of human copper-transporting ATPases. Physiol Rev. 87:1011–46.

[24]. Madsen, E., Gitlin, JD. (2007) Copper and iron disorders of the brain. Annu Rev Neurosci. 30:317–37.

[25]. Manczak M, Anekonda TS, Henson E, Park BS, Quinn J, Reddy PH (2006) Mitochondria are a direct site of A accumulation in Alzheimer's disease neurons: implications for free radical generation and oxidative damage in disease progression Hum. Mol. Gen. 15:1437-1449.

[26]. Martin LJ, Pardo CA, Cork LC, Price DL (1994) Synaptic pathology and glial responses to neuronal injury precede the formation of senile plaques and amyloid deposits in the aging cerebral cortex Am. J. Pathol. 145:1358-1381.

[27]. Mattson MP, Cheng B, Davis D, Bryant K, Lieberburg I, Rydel RE (1993) -Amyloid peptides destabilize calcium homeostasis and render human cortical neurons vulnerable to excitotoxicity J. Neurosci. 12:376-389.

[28]. Nazem, A., Mansoori, G.A. (2011) Nanotechnology for Alzheimer's disease detection and treatment. Insciences J. Nanotechnology. 1(4):169-193. doi:10.5640/insc.0104169

[29]. Onyango IG, Lu J, Rodova M, Lezi E, Crafter AB, Swerdlow RH (2010) Regulation of neuron mitochondrial biogenesis and relevance to brain health. Biochim Biophys Acta. 1802(1):228-34.

[30]. Rapoport M, Dawson HN, Binder LI, Vitek MP, Ferreira A (2002) Tau is essential to -amyloid-induced neurotoxicity Proc. Natl. Acad. Sci. USA 99:6364-6369.

[31]. Reddy PH, Beal MF (2008) Amyloid beta, mitochondrial dysfunction and synaptic damage: Implications for cognitive decline in aging and Alzheimer's disease Trends Mol. Med. 14:45-53.

[32]. Selkoe DJ (2002) Alzheimer's disease is a synaptic failure Science 298:789-791.

[33]. Skaat, H., Margel, S. (2009) Synthesis of fluorescent-maghemite nanoparticles as multimodal imaging agents for amyloid-beta fibrils detection and removal by a magnetic field. Biochem Biophys Res Commun. 386(4):645-9.

[34]. Smith, MA. (2006) Oxidative stress and iron imbalance in Alzheimer disease: how rust became the fuss! J Alzheimers Dis. 9(3 Suppl):305-8.

[35]. Swerdlow RH, Khan SM (2004) A mitochondrial cascade hypothesis for sporadic Alzheimer's disease. Med Hypotheses 63(1):8-20.

[36]. Swerdlow RH, Khan SM (2009) The Alzheimer's disease mitochondrial cascade hypothesis: an update. Exp Neurol. 218(2):308-15.

[37]. Sze C-I, Bi H, Kleinschmidt-DeMasters BK, Filley CM, Martin LJ (2001) N-Methyl-Daspartate receptor subunit proteins and their phosphorylation status are altered selectively in Alzheimer's disease J. Neurol. Sci. 182:151-159.

[38]. Sze C-I, Troncoso JC, Kawas C, Mouton P, Price DL, Martin LJ (1997) Loss of the presynaptic vesicle protein synaptophysin in hippocampus correlates with cognitive decline in Alzheimer's disease J. Neuropathol. Exp. Neurol. 56:933-994.

[39]. Taylor, JP., Hardy, J., Fischbeck, KH. (2002) Toxic proteins in neurodegenerative disease. Science Mag. 296:1991–5.

[40]. Terry RD, Masliah E, Salmon DP, Butters N, DeTeresa R, Hill R, Hansen LA, Katsman R (1991) Physical basis of cognitive alterations in Alzheimer's disease: synapse loss is the major correlate of cognitive impairment Ann. Neurol. 30:572-580.

[41]. Wang X, Su B, Siedlak SL, Moreira PI, Fujioka H, Wang Y, Casadesus G, Zhu X (2008) Amyloid-beta overproduction causes abnormal mitochondrial dynamics via differential modulation of mitochondrial fission/fusion proteins. Proc Natl Acad Sci USA 105(49):19318-23.

[42]. Yankner BA, Dawes LR, Fisher S, Villa-Komaroff L, Oster-Granite ML, Neve RL (1989) Neurotoxicity of a fragment of the amyloid precursor associated with Alzheimer's disease Science 245:417-420.

[43]. Younkin SG (1995) Evidence that Abeta 42 is the real culprit in Alzheimer's disease Ann. Neurol. 37:287-288.

[44]. Zucconi, GG., Cipriani, S., Scattoni, R., Balgkouranidou, I., Hawkins, DP., Ragnarsdottir, KV. (2007) Copper deficiency elicits glial and neuronal response typical of neurodegenerative disorders. Neuropathol Appl Neurobiol. 33:212–25.